\documentclass[12pt,tightenlines,eqsecnum,floats,aps,amsmath,amssymb,nofootinbib,prd,showpacs]{revtex4}

\usepackage{setspace} \usepackage{amsmath,amssymb,amsfonts,amsthm}
\usepackage{graphicx}

\newcommand{\be}{\begin{equation}}
\newcommand{\ee}{\end{equation}}
\newcommand{\beq}{\begin{eqnarray}}
\newcommand{\eeq}{\end{eqnarray}}

\def\lsim{\hbox{ \raise.35ex\rlap{$<$}\lower.6ex\hbox{$\sim$}\ }}
\def\gsim{\hbox{ \raise.35ex\rlap{$>$}\lower.6ex\hbox{$\sim$}\ }} 

\usepackage{bbold}

\begin{document}
\title{Lattice Refining LQC
and the Matter Hamiltonian} 
\author{William Nelson and Mairi Sakellariadou} 
\affiliation{King's College London, Department of Physics, Strand WC2R
2LS, London, U.K.}

\begin{abstract}
\vspace{.2cm}
\noindent
In the context of loop quantum cosmology, we parametrise the lattice
refinement by a parameter, $A$, and the matter Hamiltonian by a
parameter, $\delta$. We then solve the Hamiltonian constraint for both
a self-adjoint, and a non-self-adjoint Hamiltonian operator.
Demanding that the solutions for the wave-functions obey certain physical
restrictions, we impose constraints on the two-dimensional, $(A,\delta)$,
parameter space, thereby restricting the types of matter content that
can be supported by a particular lattice refinement model.
\end{abstract}
 

\maketitle

\section{Introduction}
Loop quantum gravity is a canonical quantisation of general relativity
based on a Hamiltonian formulation with basic variables the
connection, which basically carries information about curvature, and
the triad, which encodes information about the spatial
geometry. Reducing the dynamical variables of the full theory to
homogeneous and isotropic models, one gets loop quantum
cosmology~\cite{Ashtekar:2003hd}, which is not a field theory.

The fundamental variables of loop quantum cosmology are the holonomies
of the SU(2) connection, $A^i_a$, ($i$ refers to the Lie algebra index
and $a$ is a spatial index with $a$ and $i$ taking values 1,2,3) along
a given edge, and the corresponding conjugate momentum, which is the
flux of the densitised triad, $E_i^a$, through a two-surface.
Assuming spatially flat, homogeneous and isotropic models, the
connection is given by a multiple of the basis one forms, and the
triad is obtained from the determinant of the fiducial flat metric,
$^0q_{ab}$, which defines the volume, $V_0$, of the elementary cell,
${\cal V}$.  All integrations are performed over the fiducial
(elementary) cell, ${\cal V}$.

To proceed with the quantisation procedure, one has first to construct
the Hamiltonian operator.  Dynamics are then determined by the
Hamiltonian constraint. We emphasise that while in the full theory
there are an infinite number of constraints, in the reduced
homogeneous and isotropic case there is only one integrated
Hamiltonian constraint. Matter is introduced by adding the actions of
matter components to the gravitational action. Thus, one just adds the
matter contribution to the Hamiltonian constraint. One then obtains
difference equations, analogous to the differential Wheller-DeWitt
equations.

Phenomenological reasons~\cite{Vandersloot
PhD,Bojowald:2007ra,Nelson:2007wj} require the parameter appearing in
the regularisation of the Hamiltonian constraint not to be constant.
Considering an underlying lattice which is being refined during
dynamical changes of the volume, one allows the number of vertices on
the closed loop making up the holonomies to vary dynamically. One has
then to implement this requirement in the quantisation procedure of
the Hamiltonian constraint.

We parametrise the lattice refinement and the matter Hamiltonian by
introducing two parameters, $A$ and $\delta$,
respectively. Considering the Hamiltonian operator in the
self-adjoint and the non-self-adjoint cases, we solve the constraint
equation for both fixed and varying lattices. Demanding that the
solutions should satisfy certain physical requirements, we impose
constraints on the two-dimensional, $(A,\delta)$, parameter space.

\section{Constant lattice}
The gravitational part of the Hamiltonian operator, $\hat{\mathcal
C}_{\rm grav}$, can be written in terms of SU(2) holonomies,
$\hat{h}_i$, and the triad component, $p$, determining in the flat ($k=0$)
model the physical volume of the fiducial cell, ${\cal V}$,
as~\cite{Ashtekar:2006wn,Vandersloot PhD}
\be
\label{eq:ham_g1} 
\hat{\mathcal C}_{\rm grav} = \frac{2i}{\kappa^2 \hbar \gamma^3
\mu_0^3}{\rm tr} \sum_{ i j k} \epsilon^{ijk} \left(
\hat{h}_i^{(\mu_0)}\hat{h}_j^{(\mu_0)}\hat{h}i^{(\mu_0)-1}
\hat{h}_j^{(\mu_0)-1} \hat{h}_k^{(\mu_0)} \left[
\hat{h}_k^{(\mu_0)-1},\hat{V} \right]\right) {\rm sgn}(\hat p)~,
\ee 
where $\kappa = 8\pi G$, and $\hat V=\widehat{|p|^{3/2}}$ denotes the
volume operator. Note that we use the irreducible representation,
$J=1/2$, since in this case the Hamiltonian constraint is free of
ill-behaving spurious solutions~\cite{Vandersloot PhD,Perez:2005fn}.
The holonomy, $h_i^{(\mu_0)}$, along the edge parallel to the $i^{\rm
th}$ basis vector, of length $\mu_0 V_0^{1/3}$ with respect to the
fiducial metric, is~\cite{Ashtekar:2006wn}
\be
\hat h_i^{(\mu_0)} = \widehat{\cos \left( \frac{\mu_0 c}{2} 
\right)}{\mathbb 1}  +2\widehat{ \sin
\left( \frac{ \mu_0 c}{2} \right)}\tau_i~,
\ee
where ${\mathbb 1}$ is the identity $2\times 2$ matrix and
$\tau_i=-i\sigma_i/2$ is a basis in the Lie algebra SU(2) satisfying
the relation
$$\tau_i\tau_j=(1/2)\epsilon_{ijk}\tau^k-(1/4)\delta_{ij}~,$$ with
$\sigma_{\rm i}$ the Pauli matrices.  The pair $(c,p)$ denotes the
coordinates of the two-dimensional gravitational phase-space. The
triad component $p$ determines the physical volume of the fiducial
cell and the connection component determines the rate of change of the
physical edge length of the fiducial cell. They are related through
\be
\{c,p\}=\frac{\kappa\gamma}{3}~,
\ee
with $\gamma$ the Barbero-Immirzi parameter representing a quantum
ambiguity parameter of the theory.

The action of the operator $\widehat{\exp[i(\mu_0c/2)]}$ on the basis
states, $|\mu \rangle$, with
$$\hat{p}|\mu\rangle = \left( \kappa \gamma \hbar |\mu |/6 \right) |
\mu \rangle~,$$ where $\mu$ (a real number) stands for the eigenstates
of $\hat p$, satisfying the orthonormality relation
$$\langle\mu_1|\mu_2\rangle=\delta_{\mu_1,\mu_2}~,$$ reads
\be
\label{eq:exp}
\widehat{\exp \left[ \frac{i\mu_0}{2} c \right]} | \mu \rangle = 
\exp \left[ \mu_0 \frac{{\rm d}}{{\rm d} \mu}
\right] | \mu \rangle = | \mu+\mu_0 \rangle~;
\ee
$\mu_0$ is any real number.  

The action of the holonomies, $\hat{h}_i^{(\mu_0)}$, of the
gravitational connection, on the basis states is given
by~\cite{Ashtekar:2006wn}
\be
\label{eq:hol1}
\hat{h}_i^{(\mu_0)}| \mu \rangle = \left(\widehat{\rm cs} {\mathbb 1} 
+ 2 \widehat{\rm sn}\tau_{\rm i} \right) | \mu
\rangle~,
\ee
where,
\beq
\label{defin}
 \widehat{\rm cs} |\mu\rangle \equiv \widehat{\cos (\mu_0 c/2)} | \mu
\rangle &=& \left[ \ | \mu+\mu_0\rangle + | \mu -\mu_0\rangle \
\right]/2~, \nonumber \\ \widehat{\rm sn} |\mu\rangle \equiv
\widehat{\sin (\mu_0 c/2)} | \mu \rangle &=& \left[ \ |
\mu+\mu_0\rangle - | \mu -\mu_0\rangle \ \right]/(2i)~.  \eeq
Thus,
\begin{eqnarray}
\label{eq1}
&& \hat{h}_i^{(\mu_0)} \hat{h}_j^{(\mu_0)}\hat{h}^{(\mu_0)-1}_i
\hat{h}^{(\mu_0)-1}_j|\mu\rangle \nonumber\\ &&~~= \left[ \left(
\widehat{\rm cs}^4 - \widehat{\rm sn}^4 \right) {\mathbb 1} + 2\left(
{\mathbb 1} - 4 \tau_j \tau_i\right) \widehat{\rm cs}^2\widehat{\rm
sn}^2 +4\left( \tau_i - \tau_i \right) {\mathbb 1}\widehat{\rm cs}\
\widehat{\rm sn}^3 \right] | \mu \rangle~,
\end{eqnarray}
and
\begin{eqnarray}
\label{eq2}
&&\hat{h}_i^{(\mu_0)} \left[ \hat{h}_i^{(\mu_0)-1}, \hat{V} \right] |
\mu \rangle \nonumber\\ &&~~~~~~= \left( \hat{V}- \widehat{\rm
cs}\hat{V}\widehat{\rm cs} - \widehat{\rm sn}\hat{V}\widehat{\rm sn}
\right) {\mathbb 1} |\mu \rangle + 2\tau_i \left( \widehat{\rm
cs}\hat{V}\widehat{\rm sn} - \widehat{\rm sn}\hat{V} \widehat{\rm
cs} \right) | \mu \rangle~.
\end{eqnarray}
Substituting  Eqs.~(\ref{eq1}) and (\ref{eq2}) into Eq.~(\ref{eq:ham_g1}) 
we obtain
\be
\hat{\mathcal C}_{\rm grav} | \mu\rangle = \frac{48 i}{\kappa^2 \hbar
\gamma^3 \mu_0^3} \widehat{\rm cs}^2 \widehat{\rm sn}^2\left(
\widehat{\rm sn}\hat{V}\widehat{\rm cs} -\widehat{\rm
cs}\hat{V}\widehat{\rm sn} \right) | \mu \rangle~.
\ee
Using Eq.~(\ref{defin}) we recover the known expression for the action
of the gravitational part of the Hamiltonian constraint,
namely\footnote{Being interested in the large scale behaviour of the
loop quantum cosmology equations, we neglect the sign ambiguity that
arises from the two different orientations of the triad.}
\be
\label{eq:ham_g_NSA}
 \hat{\mathcal C}_{\rm grav} | \mu \rangle= \frac{1}{4}
\left(\frac{\hbar}{6\kappa \gamma^3}\right)^{1/2} \mu_0^{-3} S(\mu)
\left[ | \mu+4\mu_0\rangle -2|\mu\rangle +|\mu-4\mu_0\rangle \right]~,
\ee
where $S(\mu)$ is defined by
\begin{equation} 
S(\mu)= \left|\mu+\mu_0\right|^{3/2}-\left|\mu-\mu_0\right|^{3/2}~.
\end{equation}
To make the Hamiltonian operator self-adjoint we simply define 
\be
\hat{\mathcal H}_{\rm grav} = \frac{1}{2}\left(
\hat{\mathcal C}_{\rm grav} + \hat{\mathcal C}_{\rm grav}^\dagger \right)~,
\ee 
which acts on the basis states as
\beq
\hat{\mathcal H}_{\rm grav} |\mu\rangle &=& \frac{1}{8} \left(
\frac{\hbar}{6\kappa \gamma^3} \right)^{1/2} \mu_0^{-3} \Biggl(\Bigl[
S(\mu)+S(\mu+4\mu_0)\Bigr]|\mu +4\mu_0\rangle\nonumber \\ && 
 -4S(\mu)|\mu\rangle +\Bigl[ S(\mu) + S(\mu-4\mu_0)\Bigr]
|\mu -4\mu_0\rangle\Biggr)~.
\eeq
Taking the continuum limit ($\mu \gg \mu_0$) of the Hamiltonian
constraint equation
\be
\hat{\mathcal H}_{\rm grav}|\Psi\rangle= -\hat{\mathcal
H}_{\phi}|\Psi\rangle~,
\ee 
and expanding the general state $|\Psi\rangle$ in the kinematical
Hilbert space in terms of the basis states, $|\mu \rangle$, as
\be\label{eq:expand}
|\Psi\rangle = \sum_{\mu} \Psi_\mu (\phi)|\mu \rangle~,
\ee  
where the coefficients $\Psi_\mu$ are not continuous with respect to
$\mu$ and the dependence of the coefficients on $\phi$ represents the
matter degrees of freedom, we get
\beq
 -{\mathcal H}_\phi \Psi_\mu &=& \frac{3}{8}
\left(\frac{\hbar}{6\kappa \gamma^3}\right)^{1/2} \mu_0^{-2}\mu^{1/2}
\Biggl[ 2\Bigl( \Psi_{\mu+4\mu_0} - 2\Psi_\mu + \Psi_{\mu-4\mu_0}
\Bigr) \nonumber \\ && -\frac{2\mu_0}{\mu} \Bigl(\Psi_{\mu-4\mu_0}
-\Psi_{\mu+4\mu_0} \Bigr) - \frac{2 \mu_0^2}{\mu^2} \Bigl(
\Psi_{\mu-4\mu_0} + \Psi_{\mu+4\mu_0} \Bigr) \nonumber \\
&&-\frac{\mu_0^2}{12 \mu^2} \Bigl( \Psi_{\mu+4\mu_0}-2\Psi_\mu +
\Psi_{\mu-4\mu_0} \Bigr) + {\mathcal O}\left( \mu_0^3\right) \Biggr]~,
\eeq
where $\hat{\mathcal H}_{\phi}|\Psi\rangle= {\mathcal H}_{\phi} |\Psi
\rangle$ is assumed to act diagonally on the basis states
$|\mu\rangle$. 
We note that the kinematical inner product of the general states reads
\be
\langle \Psi|\Psi'\rangle =\sum_{\mu}\bar\Psi_\mu\Psi'_\mu~.
\ee 
with the requirement that a state in the kinematical Hilbert space
must have a finite kinematical norm.  The basis states $|\mu\rangle$
are eigenstates of the volume operator and while the eigenvalues $\mu$
are valued on the whole real line, the states are normalisable with
respect to the kinematical inner product~\cite{Vandersloot PhD}.

Assuming that the wave-function does not vary much on scales smaller
than $4\mu_0$ (known as {\it
pre-classicality}~\cite{Bojowald:2002ny}), one can approximate
$\Psi_\mu(\phi)$ as $\Psi_\mu(\phi) \approx \Psi(\mu,\phi)$.  Then
using Taylor expansion the constraint equation reduces to the
Wheeler-DeWitt equation~\cite{ABL}
\be
\label{eq:WdW_fixed}
-{\mathcal H}_{\phi}\Psi(\mu,\phi) =
6\left(\frac{\hbar}{6\kappa\gamma^3}\right)^{1/2} \left[
\frac{\partial^2}{\partial\mu^2} \left(\mu^{1/2}\Psi(\mu,\phi)\right)
+ \mu^{1/2}\frac{\partial^2 \Psi(\mu,\phi)}{\partial \mu^2}+ {\mathcal
O}\left(\mu_0\right) \right]~;
\ee
we have re-introduced the dependence on the matter degrees of freedom
$\phi$. The quantum constraint, which is a difference rather than a
differential equation, constraints the coefficients $\Psi(\mu,\phi)$ to
ensure that $|\Psi\rangle$ is a physical state.  

The non-self-adjoint version of the operator produces a different
factor ordering, namely
\be
\label{eq:WdW_fixed_NSA}
-{\mathcal H}_{\phi}\Psi(\mu,\phi) =
12\left(\frac{\hbar}{6\kappa\gamma^3}\right)^{1/2} \left[
\frac{\partial^2 }{\partial \mu^2}\left(\mu^{1/2}\Psi(\mu,\phi)\right)
+ {\mathcal O}\left(\mu_0\right) \right]~,
\ee
which affects the conditions on normalisability to be discussed later.

\section{Lattice refinement}
The case of a dynamically altering holonomy length scale,
$\tilde{\mu}(\mu)$, is required for several phenomenological
reasons~\cite{Vandersloot PhD,Bojowald:2007ra,Nelson:2007wj}.
However, this is not just a naive substitution, $\mu_0 \rightarrow
\tilde{\mu}(\mu)$, in the previous equations.  One can immediately
realise that this would lead to difficulties, since there would be
extra terms arising in Eq.~(\ref{eq:WdW_fixed}) as a result of the
dynamics of the underlying grid.

To derive the correct constraint equation we need to introduce the
varying length scale into the definition of the
holonomies~\cite{Ashtekar:2006wn}
\be
\hat{h}_{\rm i}=\widehat{\exp\left[\frac{-i\sigma_{\rm i}}{2}
\tilde{\mu c}\right]}~,
\ee
where the reader should keep in mind that $\tilde{\mu}$ depends on
$\mu$.  Geometric considerations~\cite{Ashtekar:2006wn} imply that, after
quantising, 
\be
\widehat{\exp\left[ \frac{-i\sigma_{\rm i}}{2} \tilde{\mu}c\right]} |
\Psi(\mu,\phi)\rangle = \exp\left[ \tilde{\mu}\frac{\partial}{\partial
\mu} \right] | \Psi(\mu,\phi)\rangle~.
\ee
This however is no longer a simple shift operator, since $\tilde{\mu}$
is a function of $\mu$.  Consider changing the representation from
$\mu$ to
\be
\nu = \tilde\mu_0 \int \frac{{\rm d}\mu}{\tilde{\mu}(\mu)}~,
\ee
where $\tilde\mu_0$ is a constant. In this representation we have
\be
\hat{h}_{\rm i} |\nu \rangle = \exp \left[\tilde{\mu}(\mu) \frac{\rm
d}{{\rm d} \mu}\right] | \nu \rangle =\exp \left[ \tilde\mu_0
\frac{\rm d}{{\rm d} \nu}\right]|\nu \rangle=|\nu+\tilde\mu_0\rangle~.
\ee
We can then proceed as before and define
\beq
\widehat{\rm sn}|\nu\rangle &\equiv&
\widehat{\sin\left[\frac{\tilde{\mu}c}{2}\right]} |\nu\rangle =
\frac{1}{2i} \Bigl[ | \nu + \tilde{\mu}_0\rangle -|\nu - \tilde{
\mu}_0\rangle \Bigr]~, \nonumber\\
\widehat{\rm cs}|\nu\rangle &\equiv& \widehat{\cos\left[\frac{
\tilde{\mu}c}{2}\right]} |\nu\rangle = \frac{1}{2} \Bigl[ | \nu
+ \tilde{\mu}_0\rangle +|\nu - \tilde{\mu}_0\rangle \Bigr]
\nonumber~.
\eeq
There is however a problem in defining the volume eigenvalue, since
this requires an explicit relation between $\nu$ and $\mu$ given by
$\tilde{\mu}$. Assuming
\be\tilde{\mu}=\mu_0 \mu^A~,
\ee 
one has
\be
\label{def:nu}
\nu=\frac{\tilde\mu_0\mu^{1-A}}{\mu_0(1-A)}
\ee 
(up to a constant that can be set equal to $0$), leading to
\beq
\hat{V}|\nu\rangle &=&\left(\frac{\kappa\gamma\hbar}{6}\right)^{3/2}
\mu^{3/2}|\nu\rangle\nonumber\\ &=& \left(\frac{\kappa \gamma
\hbar}{6}\right)^{3/2} \left[ \frac{\mu_0\left(1-A\right)}{
\tilde\mu_0}\right]^{3/2/(1-A)} \nu^{3/2/(1-A)} |\nu\rangle~.
\eeq

\subsection{Non-self-adjoint case}
Let us calculate the action of Eq.~(\ref{eq:ham_g1}) on the basis
state $|\nu\rangle$:
\be
\hat{\mathcal C}_{\rm grav}|\nu\rangle = \frac{1}{4\mu_0^3}\left(
\frac{\hbar}{6\kappa\gamma^3}\right)^{1/2}\left(\alpha\nu\right)^{3A/(A-1)}
S(\nu) \Bigl(|\nu+4\tilde\mu_0\rangle -2|\nu\rangle +|\nu
-4\tilde\mu_0\rangle \Bigr)~,
\ee
where 
\be
\alpha=\mu_0(1-A)/\tilde\mu_0~,
\ee 
and $S(\nu)$ is defined by
\be
S(\nu) = \Bigl[\left(\nu+\tilde\mu_0\right)\alpha\Bigr]^{3/2/(1-A)}
-\Bigl[\left(\nu-\tilde\mu_0\right) \alpha\Bigr]^{3/2/(1-A)}~.
\ee
One can easily check that $A=0$ reproduces Eq.~(\ref{eq:ham_g_NSA}),
if it is taken that $\tilde{\mu}_0=\mu_0$. After a long but straight
forward expansion in the $\nu\gg \tilde\mu_0$ limit and under the
assumption of pre-classicality, one finds
\beq
\hat{\mathcal C}_{\rm grav}|\Psi(\nu,\phi)\rangle&=& \sum_{\nu}
12(1-A)^2\left(\frac{\hbar}{6\kappa\gamma^3}\right)^{1/2}
\alpha^{-3/2\left(1-A\right)}\ 
\nu^{\left(1-4A\right)/2\left(1-A\right)} \nonumber \\
&&\times \Biggl[ \frac{\partial^2
\Psi(\nu,\phi)}{\partial \nu^2} +\frac{1-4A}{(1-A)\nu}
\frac{\partial \Psi(\nu,\phi)}{\partial \nu} +
\frac{\left(1+2A\right)\left( 4A-1\right)}{4\left(1-A\right)^2 \nu^2}
\Psi(\nu,\phi) \nonumber \\
&&\ \ \ \ \ + {\mathcal O}\left(\tilde\mu_0\right)\Biggr]|\nu\rangle~.
\eeq
For a non-self-adjoint Hamiltonian operator, the Hamiltonian
constraint equation reads
\be
\frac{\partial^2 \Psi(\nu,\phi)}{\partial \nu^2} + \frac{B}{\nu}
\frac{\partial\Psi(\nu,\phi)}{ \partial \nu} + \frac{C}{\nu^2}
\Psi(\nu,\phi) +\beta {\mathcal H}_{\phi} \nu^{-B/2}\Psi(\nu,\phi)
+{\mathcal O}\left(\tilde\mu_0\right)=0~,
\ee
where
\beq
B&=&\frac{1-4A}{(1-A)} \nonumber \\ C&=&\frac{
\left(1+2A\right)\left(4A-1\right)}{4\left(1-A\right)^2}\nonumber\\
\beta&=&\frac{\alpha^{3/2/(1-A)}}{12(1-A)^2}\left(\frac{6\kappa\gamma^3}
{\hbar}\right)^{1/2}
~.
\label{eq:beta}
\eeq 
We note that for a fixed lattice, $A, B, C$ and $\beta$ are given by
\be
A=0 \ \ ,\ \ B=1 \ \ , \ \ C=-1/4 \ \ , \ \ \beta =
\left(6\kappa\gamma^3/\hbar\right)^{1/2} \left(
\mu_0/\tilde\mu_0\right)^{3/2}/12~.
\ee 
Considering lattice refinement in the case of a non-self-adjoint Hamiltonian
operator, one has $\nu=\tilde\mu_0\mu/\mu_0$. Thus, by keeping
$\tilde\mu_0$ general all we have done is to re-scale $\mu$. Setting
$\mu_0=\tilde\mu_0$ we get back Eq.~(\ref{eq:WdW_fixed_NSA}).

The specific lattice refinement
$A=-1/2$~\cite{Ashtekar:2006wn,Bojowald:2007ra} is clearly a
particularly fortitious choice as it results in $C=0$. Notice however
that choosing $A=1/4$ results in a further simplification, leading to
\be
\frac{\partial^2 \Psi(\nu,\phi)}{\partial\nu^2} + \frac{1}{12}
\left(\frac{6\kappa \gamma^3}{\hbar} \right)^{1/2} {\mathcal H}_{\phi}
\Psi(\nu,\phi) =0~,
\ee
assuming $\tilde\mu_0=\mu_0$. This unphysical lattice refinement
choice results in dynamics that are well approximated by the
Wheeler-DeWitt equation in the large scale limit (for slowly varying
wave-functions). This remark highlights the importance of
understanding the origin of the lattice refinement in the full theory.
Unfortunately, at present there is little theoretical reason for
discounting such physically unacceptable scenarios, and one must 
rely on his/her phenomenological intuition.

\subsection{Self-adjoint case}
Let us repeat the above procedure for the case of a self-adjoint
Hamiltonian operator, 
$$\hat{\mathcal H}_{\rm grav} = (\hat{\mathcal C}_{\rm grav} +
\hat{\mathcal C}_{\rm grav}^{\dagger})/2~.$$
Acting on the state $|\nu\rangle$ one has
\beq
\hat{\mathcal H}_{\rm grav}|\nu\rangle &=& \frac{1}{8\mu_0^3}\left(
\frac{\hbar}{6\kappa\gamma^3}\right)^{1/2} \left(\alpha
\nu\right)^{3A/(A-1)} \Biggl[ \Bigl\{ S(\nu + 4\tilde\mu_0) + S(\nu)
\Bigr\} |\nu + 4\tilde\mu_0\rangle \nonumber \\ &&-4S(\nu) + \Bigl\{
S(\nu-4\tilde\mu_0)+S(\nu)\Bigr\}|\nu-4\tilde\mu_0\rangle \Biggr]~.
\eeq
Expanding $\hat{\mathcal H}_{\rm grav}|\Psi\rangle = -\hat{\mathcal
H}_{\phi}|\Psi \rangle$, one obtains
\be\label{eq:ham_SA_var}
\frac{\partial^2 \Psi(\nu,\phi)}{\partial \nu^2} +
\frac{\tilde{B}}{2\nu}\frac{\partial \Psi(\nu,\phi)}{ \partial \nu} +
\tilde{C}\nu^{-2}\Psi(\nu,\phi)+ \beta{\mathcal H}_{\phi}
\nu^{-B/2}\Psi(\nu,\phi) + {\mathcal O}\left(\tilde\mu_0 \right) =
0 ~,
\ee 
where $B$ and $\beta$ are given by Eq.~(\ref{eq:beta}a) and
Eq.~(\ref{eq:beta}c), respectively, and
\beq
 \tilde{B}&=&\frac{1-10A}{1-A}~, \nonumber\\
 \tilde{C}&=&\frac{(1+2A)(4A-1)+12A(2A-1)}{8(1-A)^2}~.
\eeq
As expected, setting $A=0$ and $\tilde\mu_0=\mu_0$ gives back
Eq.~(\ref{eq:WdW_fixed}). Once again we see that the choice
$A=-1/2$~\cite{Ashtekar:2006wn,Bojowald:2007ra} produces a particular
simplification.

\subsection{Physical sector}\label{sec:norm}

In general, not all solutions to the quantum constraint equation,
Eq.~(\ref{eq:ham_SA_var}) in the case of a self-adjoint Hamiltonian
operator, are normalisable with respect to the physical inner
product~\cite{Ashtekar:2006wn,Ashtekar:1994kv}. In what follows, we
are only interested in physical states. The physical Hilbert space
consists of solutions to the quantum constraint equation which have
finite norm with respect to the physical inner product.  The inner
product on physical states can be obtained by requiring that real
classical observables be represented on the physical Hilbert space by
self-adjoint operators~\cite{Ashtekar:1994kv}.  The physical inner
product has been calculated~\cite{Ashtekar:2006wn,Ashtekar:1994kv} if
the only matter source is a massless scalar field. Following the same
procedure, we will compute the inner product of physical states for
the model we are considering here.

The (total) Wheeler-DeWitt constraint equation reads 
\be 
\left(\hat{\cal H}_{\rm grav} +\hat{\cal H}_\phi\right)\Psi = 0~.  
\ee
Since we are interested in the large scale limit, we approximate the
matter Hamiltonian, $\hat{\cal H}_\phi$, with $\hat{\cal H}_\phi=
\hat{\nu}^\delta\hat{\epsilon}\left(\phi\right)$ (the reader is
referred to the next Section). Thus,
\be
 \hat{\epsilon}\left(\phi\right)\Psi \equiv \epsilon\left(\phi\right)\Psi =
  -\nu^{-\delta}\hat{\cal H}_{\rm grav} \Psi~.
\ee
In the classical theory, $\epsilon\left(\phi\right)$ is a Dirac
observable since it is a constant of
motion~\cite{Ashtekar:2006wn}. Even though $\nu(\phi)$ is not a
constant of motion, assuming that $\nu(\phi)$ is a monotomic function
(with respect to $\phi$), then $\nu(\phi_0)$ is a Dirac observable for
any fixed $\phi_0$~\cite{Ashtekar:2006wn}. Modulo an overall scaling,
the unique inner product which makes these operators self-adjoint
is~\cite{Ashtekar:2006wn}
\be
\label{eq:norm} 
\langle \Psi_1 | \Psi_2\rangle_{\rm phys} = \int_{\phi=\phi_0}d\nu
|\nu |^\delta \overline{\Psi}_1\Psi_2~.  
\ee
The finite norm of the physical wave-functions, defined by
Eq.~(\ref{eq:norm}), is conserved, {\sl i.e.}, independent of the
choice of $\phi=\phi_0$. From Eq.~(\ref{eq:norm}) one concludes that
the solutions of the constraint are normalisable provided they decay,
on large scales, faster than $\nu^{-1/(2\delta)}$.  One arrives to the
same conclusion for the case of a constant lattice, with $\nu$
replaced by $\mu$.

It is important to note that, in general the approximation of
$\hat{\cal H}_\phi= \hat{\nu}^\delta\hat{\epsilon}\left(\phi\right)$
is only valid on the large scale $\nu$ limit, implying that the
integrand of Eq.~(\ref{eq:norm}) is only valid for $\nu \gg
1$. However, it is certainly {\it necessary} that the large scale
behaviour of the wave-functions be normalisable with respect to
Eq.~(\ref{eq:norm}). Thus, the constraint we have found is a {\it
necessary} but not {\it sufficient} condition for the wave-functions
to be considered physical.

\section{Solving the constraint equation}

To solve the constraint equation one needs to know the specific form
of $\mathcal{H}_\phi$.  In general, $\mathcal{H}_\phi$ has two terms
with different scale dependence, however since we are concerned only
with the large scale limit, one of these terms will be the dominant
one. Making this approximation, one can write
\be
\beta{\mathcal H}_{\phi}=\epsilon_\mu(\phi) \mu^{\delta_\mu} \ \ \ \
{\mbox
or} \ \ \ \
\beta{\mathcal H}_{\phi} =\epsilon_\nu(\phi) \nu^{\delta_\nu}~,
\ee 
where the functions $\epsilon_\mu$, $\epsilon_\nu$ are constant with
respect to $\mu$, $\nu$, respectively. The general analytical
solutions read
\beq
^{\rm \ non-self-adjoint,\ fixed \ lattice\ }\Psi(\mu)&=&C_1 \mbox{
\large \it{J}}_{2/(3+2\delta_\mu)} \left(
\frac{4\sqrt{\epsilon_\mu}}{3+2\delta_\mu}\ \mu^{(3+2\delta_\mu)/4}\right)
\nonumber \\ &&+C_2 \mbox{ \large \it{Y}}_{2/(3+2\delta_\mu)} \left(
\frac{4\sqrt{\epsilon_\mu}}{3+2\delta_\mu} \mu^{(3+2\delta_\mu)/4}\right)
\nonumber \\
\nonumber \\
^{\rm self-adjoint,\ fixed \ lattice\ }\Psi(\mu)&=&C_1 \mu^{1/4}
\mbox{ \large \it{J}}_{\sqrt{3}/(3+2\delta_\mu)} \left(
\frac{4\sqrt{\epsilon_\mu}}{3+2\delta_\mu}\ \mu^{(3+2\delta_\mu)/4}\right)
\nonumber \\ &&+C_2 \mu^{1/4}\mbox{ \large
\it{Y}}_{\sqrt{3}/(3+2\delta_\mu)} \left(
\frac{4\sqrt{\epsilon_\mu}}{3+2\delta_\mu} \ \mu^{(3+2\delta_\mu)/4}\right)
\nonumber \\
\nonumber \\
\label{eq:NSA_var}
^{\rm \ non-self-adjoint,\ varying \ lattice\ }\Psi(\nu)&=&C_1
\nu^{-3A/2/(A-1)} \mbox{ \large \it{J}}_{(2x)^{-1}} \left(
\frac{\sqrt{\epsilon_\nu}}{x} \nu^{x}\right) \nonumber \\ &&+C_2
\nu^{-3A/2/(A-1)} \mbox{ \large \it{Y}}_{(2x)^{-1}} \left(
\frac{\sqrt{\epsilon_\nu}}{x} \nu^{x}\right)
\nonumber \\
\nonumber \\
\label{eq:SA_var}
^{\rm self-adjoint,\ varying \ lattice\ }\Psi(\nu) &=& C_1
\nu^{(1+8A)/4/(1-A)} \mbox{ \large \it{J}}_{ x^{-1}y } \left(
\frac{\sqrt{\epsilon_\nu}}{x} \nu^{x}\right) \nonumber \\ &&+C_2
\nu^{(1+8A)/4/(1-A)} \mbox{ \large \it{Y}}_{ x^{-1}y } \left(
\frac{\sqrt{\epsilon_\nu}}{x} \nu^{x}\right)~,
\eeq
with 
\beq
x&=&\frac{2\delta_\nu(1-A)+3}{4(1-A)}~
\nonumber\\
y&=&\frac{\sqrt{3(12A+1)}}{4(1-A)}~;
\eeq
$J$ and $Y$ are Bessel functions of the first and second kind,
respectively, and $C_1, C_2$ are integration constants. Note that we
suppressed the $\phi$ dependence for clarity.  We wrote explicitly the
solutions for the non-self-adjoint, as well the self-adjoint case for
both a fixed and a varying lattice.  In particular, for the {\it
physically} justified choice
$A=-1/2$~\cite{Ashtekar:2006wn,Bojowald:2007ra} the solution of the
Hamiltonian constraint equation, in the case of a non-self-adjoint
Hamiltonian operator and a varying lattice, reads
\be
\Psi(\nu)= C_1 \nu^{-1/2} \mbox{ \large
\it{J}}_\frac{i\sqrt{15}}{3(\delta_\nu +1)} \left(
\frac{2\sqrt{\epsilon_\nu}}{\delta_\nu+1}
\nu^{(\delta_\nu+1)/2}\right) +C_2\nu^{-1/2}\mbox{\large
\it{Y}}_\frac{i\sqrt{15}}{3(\delta_\nu+1)} \left(
\frac{2\sqrt{\epsilon_\nu}}{\delta_\nu+1}\nu^{(\delta_\nu+1)/2}\right)~.
\ee
Among such solutions, we will only consider the physical ones.  This
immediately eliminates the solutions to the non-self-adjoint
Hamiltonian constraint, since it is not possible to find self-adjoint
Dirac observables for these cases.  We nevertheless find interesting
to compare the non-self-adjoint solutions to the self-adjoint ones and
we thus apply the norm defined by Eq.~(\ref{eq:norm}) to both
(self-adjoint and non-self-adjoint) sets of solutions. This is done
simply to complete the formal comparison between the two cases and it
is to be remembered that the normalisation constraint produced here is
only rigorous for the self-adjoint case.

Using the asymptotic expansions of the Bessel functions, \beq
\lim_{z\rightarrow \infty} \mbox{ \large \it{J}}_\beta (z)
&\rightarrow& \sqrt{\frac{2}{\pi z}} \cos \left( z - \frac{\beta
\pi}{2} - \frac{\pi}{4} \right) \nonumber \\ \lim_{z\rightarrow
\infty} \mbox{ \large \it{Y}}_\beta (z) &\rightarrow&
\sqrt{\frac{2}{\pi z}} \sin \left( z - \frac{\beta \pi}{2} -
\frac{\pi}{4} \right)~, \nonumber \eeq we find that the solutions
oscillate within an envelope that scales as \beq ^{\rm \
non-self-adjoint,\ fixed \ lattice\ }\Psi(\mu) &\propto&
\mu^{-(3+2\delta_\mu)/8} \nonumber\\ ^{\rm \ self-adjoint,\ fixed \
lattice\ }\Psi(\mu) &\propto& \mu^{-(1+2\delta_\mu)/8}\nonumber \\
^{\rm \ non-self-adjoint, \ varying \ lattice\ }\Psi(\nu) &\propto&
\nu^{[12A-2\delta_\nu(1-A)-3]/8/(1-A)}\nonumber \\ ^{\rm \
self-adjoint,\ varying \ lattice\ }\Psi(\nu) &\propto&
\nu^{[16A-1-2\delta_\nu(1-A)]/8/(1-A)}~.  \eeq As shown in Section
\ref{sec:norm} if the solutions are to be normalisable, $\Psi(\nu)$
must not grow faster than $\Psi(\nu) \propto \nu^{-1/(2\delta)}$,
which imposes constraints on the scale dependence of the allowed
matter component. More precisely, \beq -\frac{3}{4} <
\frac{\delta_\mu}{2}-\frac{1}{\delta_\mu} && \mbox{non-self-adjoint,\
fixed \ lattice}\nonumber\\ -\frac{1}{4} <
\frac{\delta_\mu}{2}-\frac{1}{\delta_\mu}&& \mbox{self-adjoint,\ fixed
\ lattice}\nonumber\\ \frac{12A-3}{4(1-A)} <
\frac{\delta_\nu}{2}-\frac{1}{\delta_\nu} && \mbox{non-self-adjoint,\
varying \ lattice}\nonumber\\ \frac{16A-1}{4(1-A)} <
\frac{\delta_\nu}{2}-\frac{1}{\delta_\nu} && \mbox{self-adjoint,\
varying \ lattice}~.  \eeq In the self-adjoint and non-self-adjoint
Hamiltonian operator cases with lattice refinement, the growth is
taken w.r.t.  $\nu$. If we require the semi-classical wave-functions
not to grow w.r.t. $\mu$, there is an additional constraint, namely
$A<1$, to ensure that increasing $\mu$ corresponds to increasing
$\nu$.

One should also keep in mind that since the solutions are only valid
on large scales, one must ensure that the large argument expansions of
the Bessel functions apply in this limit. The expansions are valid for
\be
 \delta_\mu>-3/2~,
\ee
in both cases of a self-adjoint and a
non-self-adjoint Hamiltonian operator considering a fixed lattice, and
for 
\be
 \delta_\nu > \frac{3}{2(A-1)}~,
\ee
in both cases of a
self-adjoint and a non-self-adjoint Hamiltonian operator considering a
varying lattice. Beyond these limits the wave-functions, on large scales 
decay like $1/y$, where the argument of the Bessel
functions is $\mu^{y}$, or $\nu^{y}$, respectively. Whilst these
wave-function may be normalisable, they lack a semi-classical
interpretation and hence would not produce classical cosmology at
large scales~\cite{Bojowald:2002xz}.  The regions where the different
wave-function coefficients are bounded are shown in Fig.~\ref{fig3} along
with the limit of the expansion.

\begin{figure}
 \begin{center}
   \input{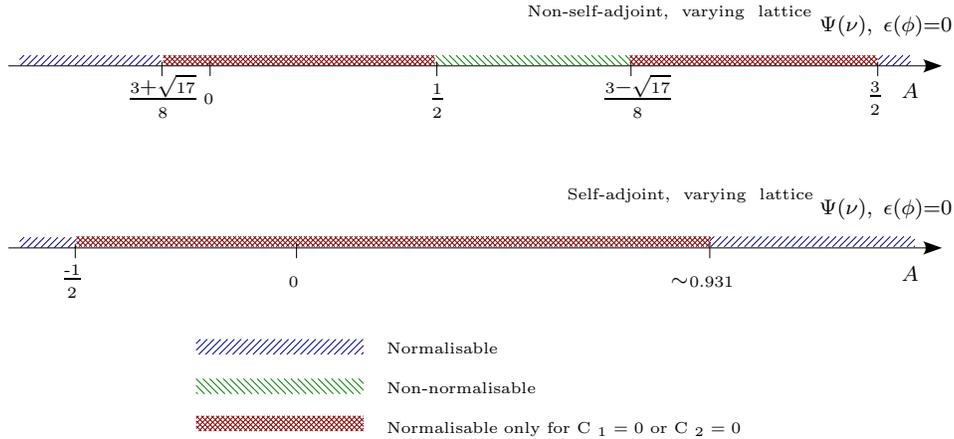}
\caption{\label{fig2} Regions in which the solutions to
the different vacuum Hamiltonian constraint equations are normalisable at large
scales. Note that $^{\rm self-adjoint,\ varying\
lattice\ } \Psi(\nu)$ is oscillatory only for $A<-1/12$.}
  \end{center}
 \end{figure}

A particularly interesting case is that of the vacuum, where
${\mathcal H}_{ \phi}=0$. This corresponds to
$\epsilon\left(\phi\right)=0$ or $\delta_\nu=-\infty$, which makes the
norm calculated in Section~\ref{sec:norm} trivial. In this case the
norm can be taken to be,
\be\label{eq:norm2}
 \langle \Psi_1 | \Psi_2\rangle = \int_{\phi=\phi_0}d\nu
\overline{\Psi}_1\Psi_2~,
\ee
and correspondingly for $\mu$. The solutions to the four constraint
equations are
\beq
^{\rm non-self-adjoint\ ,\ fixed\ lattice\
}\Psi(\mu)&=&C_1\mu^{1/2}+C_2\mu^{-1/2}\nonumber  \\  ^{\rm self-adjoint\ ,\
fixed\ lattice\ }\Psi(\mu)&=&C_1\mu^{(1+\sqrt{3})/4} + C_2
\mu^{(1-\sqrt{3})/4} \nonumber \\ ^{\rm non-self-adjoint\ ,\ varying\
lattice\
}\Psi(\nu)&=&\tilde{C}_1\nu^{(4A-1)/2}+\tilde{C_2}\nu^{(2A+1)/2} \nonumber \\
^{\rm self-adjoint\ ,\ varying\ lattice\ }\Psi(\nu)&=&\tilde{C}_1
\nu^{(8A+1+\sqrt{36A+3})/4}
+ \tilde{C}_2
\nu^{(8A+1-\sqrt{36A+3})/4}~.
\eeq
Clearly the two cases of Hamiltonian operator for a fixed lattice are
bounded only for specific choices of the integration constants,
which amounts to special initial conditions. For the lattice
refinement case however, there are several regions in which the
solutions are bounded, shown in Fig.~\ref{fig2}. It is also
worth noticing that only the self-adjoint lattice refinement case
produces oscillatory solutions (for $A<-1/12$) and hence have a simple
semi-classical dynamical interpretation~\cite{Bojowald:2002xz}.

\section{Large scale classical breakdown}

The form of the wave-functions indicates that the period of
oscillations can decrease as the scale increases, which implies that
at sufficiently large scales the assumption that the wave-functions
are pre-classical may break
down~\cite{Bojowald:2002xz,Vandersloot2005}.  This would then lead to
quantum gravity corrections at large scale (classical) physics. The
need to avoid this undesired event, was indeed one of the motivations
behind lattice refinement. However, the level of lattice refinement
necessary depends on the matter content.

\begin{figure}
\begin{center}
 \input{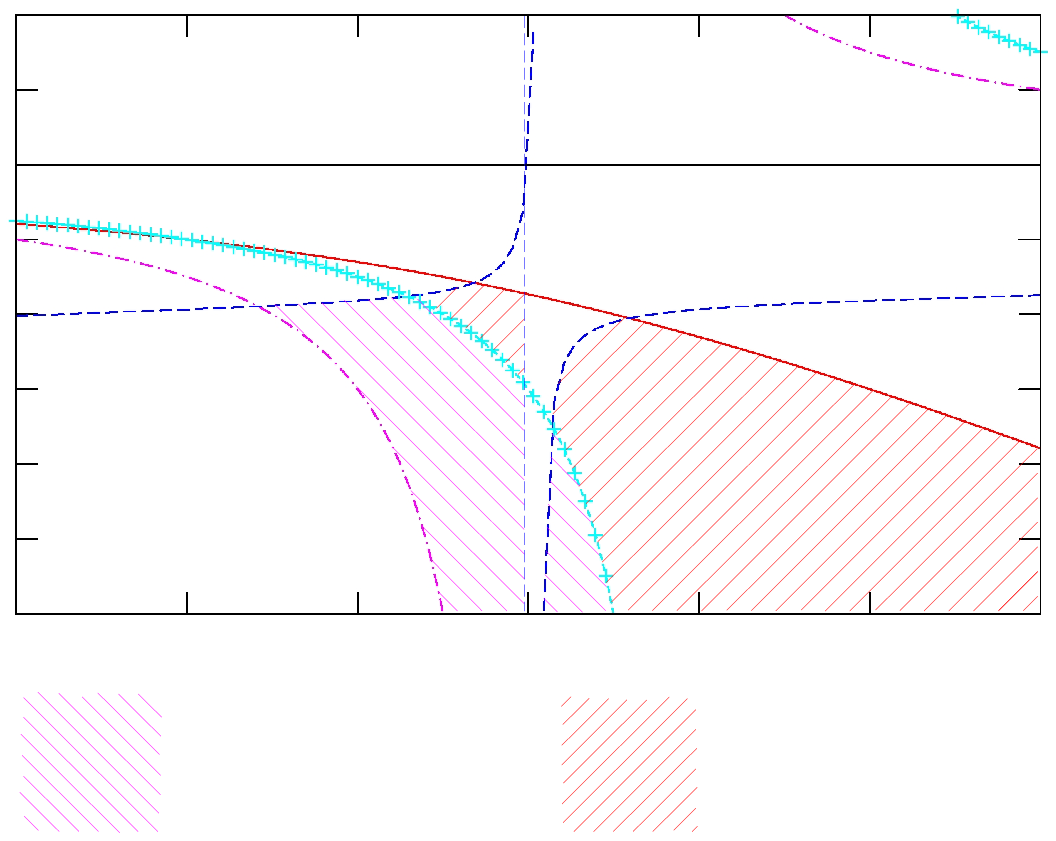}
 \caption{\label{fig3} The regions of parameter space in which the
wave-functions of the self-adjoint Hamiltonian constraint equation
with lattice refinement are physically acceptable. Notice that there
are regions (crosses) in which the Taylor expansions used to calculate
the large scale behaviour of the wave-functions are no longer valid, and
hence whilst we can say that these wave-functions decay sufficiently quickly
on large scales to be normalisable and are physical (i.e. oscillating),
we cannot be sure that there is no new large scale behaviour due to the
underlying discreteness.}
\end{center}
\end{figure}

One way to investigate this is to look at the separation between the
zeros of the wave-functions~\cite{Nelson:2007wj} (for an alternative
method see Ref.~\cite{Bojowald:2007ra}). For the non-self adjoint
Hamiltonian case we find that the $n^{th}$ zero (for large scales)
occurs at
\be
\label{eq:nu_n}
\nu_n=\left[\frac{\pi \left( 2\delta_\nu(A-a)
-3\right)}{4(A-1)\sqrt{\epsilon_\nu}}\right]^{4(A-1)/
[2\delta_\nu(A-1)-3]} \left(n + C\right)^{4(A-1)/
[2\delta_\nu(A-1)-3]}~,
\ee
where 
\be
C= \frac{1}{\pi}\tan^{-1}\left(-\frac{C_1}{C_2}\right) + \frac{(A-1)}{
2\delta_\nu(A-1)-3} \pm \frac{1}{2} + \frac{1}{4}~,
\ee 
is a constant.  Since Eq.~(\ref{eq:nu_n}) is derived from the large 
argument expansion of Eq.~(\ref{eq:NSA_var}), it is only valid for
$\delta_\nu > 3/2/(A-1)$. Using a Taylor expansion we find
\be
\label{eq:delta_nu_NSA}
\lim_{\rm large \nu} \Delta \nu_n = \frac{\pi}{\sqrt{\epsilon_\nu}}
\nu^{\frac{(4-2\delta_\nu)(1-A) -3}{4(1-A)}} + {\cal
  O}\left(\nu^{\frac{4(1-\delta_\nu)(1-A)-6}{4(1-A)}}\right)~.
\ee
Note that the Taylor expansion is valid for $\delta >
1+3/2/(A-1)$. Using $$\nu=\tilde{\mu}_0 \frac{\mu^{1-A}}{\mu_0 (1-A)}~,$$ 
we find
\be
\label{eq:delta_nu}
 \lim_{\rm large \nu}\Delta \nu_n \propto \mu_n^{(\delta_\nu-2)(A-1)/2 -3/4} +
{\mathcal O}\left( \mu_n^{ (\delta_\nu-1)(A-1) -3/2
}\right)~.
\ee
The lattice refinement will support all oscillations of the
wave-function, provided $\Delta \nu_n$ is larger than $\nu_c $, the
scale at which the underlying discreteness becomes important. The
condition for the continuum limit to be valid, is that the
wave-function must vary slowly on scales of the order
of~\cite{Nelson:2007wj}
\be
\mu_{\rm c} =4\tilde{\mu}~,
\ee
or, equivalently,
\be
\mu_{\rm c}
=4\mu_0 \mu^A~.
\ee 
Then from Eq.~(\ref{def:nu}) one gets
\be
\nu_c = \frac{\tilde{\mu}_0(4\mu^A)^{1-A}}{\mu_0^A (1-A)}~.
\label{nuc}
\ee 
Equations (\ref{eq:delta_nu}) and (\ref{nuc}) imply that lattice
refinement will be sufficient to prevent quantum corrections becoming
significant at large scales provided
\be
\label{eq:class_bd_NSA}
f(A,\delta_\nu) \equiv A^2 +\left(\frac{\delta_\nu}{2}-2\right)A
+\frac{1}{4} - \frac{\delta_\nu}{2} \geq 0~,
\ee
with further restrictions on $\epsilon_\nu(\phi)$ for the case of 
equality~\cite{Nelson:2007wj}. This is shown in Fig.~\ref{fig3}.

A similar calculation for the self-adjoint case gives
\be
\lim_{\rm large \ \nu} \Delta \nu_n = \frac{\pi}{\sqrt{\epsilon_\nu}}
\nu^{\frac{(4-2\delta_\nu)(1-A) -3}{4(1-A)}} + {\cal
 O}\left(\nu^{\frac{4(1-\delta_\nu)(1-A)-6}{4(1-A)}}\right)~,
\ee
which is precisely the same equation we had for the non-self-adjoint
case, Eq~(\ref{eq:delta_nu_NSA}). In addition, the Taylor and Bessel
expansions used are valid in the same ranges as those of the
non-self-adjoint case. This is not surprising since making the
constraint equation self-adjoint is inherently a quantum operation and
their classical limits should be the same. Thus, Fig.~\ref{fig3}
applies to self-adjoint, as well as to non-self-adjoint Hamiltonian
operators, in the lattice refinement case (albeit with a different 
constraint coming from the requirement that the coefficients be normalisable).

\begin{figure}
 \begin{center}
  \input{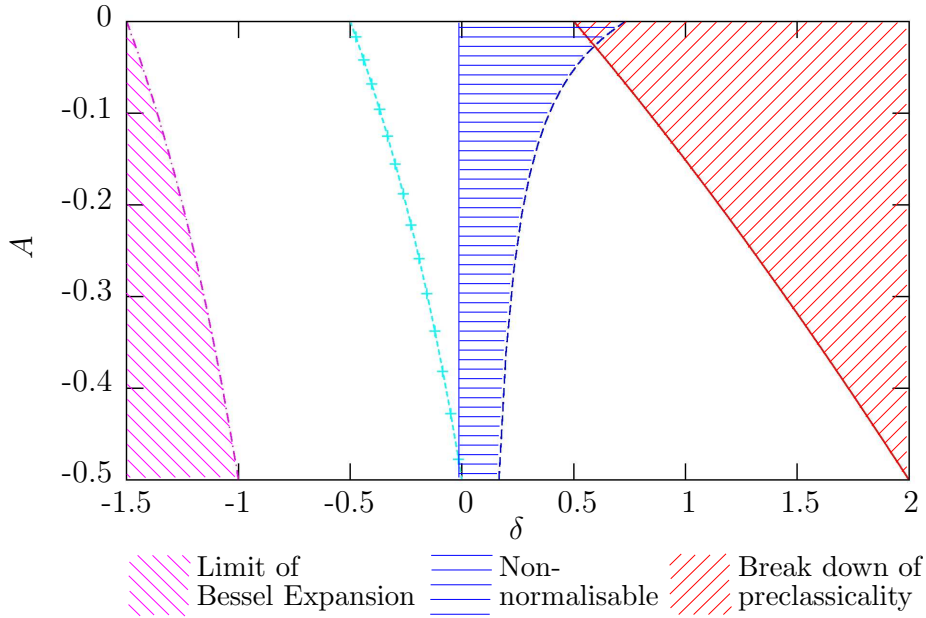}
  \caption{\label{fig6} The full loop quantum gravity theory allows
only the range $0<A<-1/2$. Within this range we can see that the
acceptable types of matter content are significantly restricted. In
addition, notice that for a varying lattice ($A \neq 0$) it is not
always possible to treat the large scale behaviour of the
wave-functions perturbatively (dashed line with crosses).}
 \end{center}
\end{figure}

By considering the underlying origins of lattice refinement, we can
further restrict the allowed range to
$0<A<-1/2$~\cite{Bojowald:2007ra}. This allows us to examine the types
of matter that cannot be supported by a particular lattice refinement
model. The relevant section of Fig.~\ref{fig3} is replotted in
Fig.~\ref{fig6}. Notice that Eq.~(\ref{def:nu}) is needed to find the scaling
behaviour of a particular matter component with respect to the scale factor,
i.e. matter scaling like $\nu^{\delta_\nu}$, scales with respect to the
scale factor as $a^{2\delta_\nu (1-A)}$.

\section{Conclusions}
We have derived, in the continuum limit, the Hamiltonian constraint of
loop quantum cosmology for a general lattice refinement scheme of the
form $\tilde{\mu}= \mu_0\mu^A$, for both the self-adjoint and
non-self-adjoint Hamiltonian operator cases. We solved the resulting
Wheeler-deWitt like equations and discussed the requirements the
solutions must satisfy in order to be physically viable. These
requirements give us constraints on the type of matter that can be
supported by a particular lattice refinement model.  We considered the
following three requirements for the wave-functions: (i) that the
coefficients of their basis expansion be normalisable, (ii) that they
have oscillating large scale solutions so as to ensure that classical
dynamics can be recovered and (iii) that they are pre-classical at
large scales.  Combining these conditions significantly constrains the
allowable region of parameter space.

In particular, for the case of a constant lattice, physical
wave-functions are produced only if ${\cal H}_\phi$ scales faster than
$a^{-1}$ and slower than $a$; an extremely severe restriction, given
that many types of matter scale beyond this range. In the most popular
lattice refinement model, $A=-1/2$, this range is extended so that
physical wave-functions are produced provided ${\cal H}_\phi$ scales
faster than $a^{-3}$ and slower than $a^6$, although it is not
possible to treat the large scale oscillations perturbatively over a
third of this range ($a^{-3} \rightarrow a^0$).

As a concrete example, an inflationary scalar field (i.e., one in
which the potential term dominates over the kinetic term in the matter
Hamiltonian), scales like $a^3$. From our general procedure it is
clear that this has a large scale breakdown of pre-classicality for
the fixed lattice case, whilst this problem is resolved for the
common, $A =-1/2$, lattice refinement case, as was shown
in~\cite{Nelson:2007wj}.  This provides a further demonstration of the
importance of modelling lattice refinement in loop quantum cosmology,
if physical results are to be produced, and it does so for a large
class of such models.

It is important to note that lattice refinement could, in principle,
be much more complicated than the power law form ($\tilde{\mu}= \mu_0
\mu^A$) used here, however even with this simplifying assumption the
qualitative behaviour of different lattice refinement models is clear.
In particular, we have shown that the continuum limit of the Hamiltonian
constraint equation is sensitive to the choice of model and that 
only a limited range of matter components can be supported within a
particular choice. This further emphasises the need to support 
effective, phenomenological lattice refinement models with a deeper
understanding of the fundamental theory.

\vskip.05truecm 
\acknowledgements 
We would like to thank the anonumous referee for his/her useful
comments regarding the normalisation condition on the physical states.
This work is partially supported by the European Union through the
Marie Curie Research and Training Network \emph{UniverseNet}
(MRTN-CT-2006-035863).

\end{document}